\documentclass[10pt]{article}
\usepackage{amsmath}
\usepackage{graphicx}
\usepackage{amssymb}
\usepackage{latexsym}
\usepackage{bm}
\newcommand{\vect}[1]{\bm{\mathrm{#1}}}
\usepackage{mathrsfs}
\usepackage[OE]{express}

\begin{document}

\title{Frequency-modulated continuous-wave LiDAR compressive depth-mapping}

\author{Daniel J. Lum\authormark{1,2,*}, Samuel H. Knarr\authormark{1,2,*} and John C. Howell\authormark{1,2,3}}

\address{
\authormark{1}Department of Physics and Astronomy, University of Rochester, Rochester, New York 14627, USA\\
\authormark{2}Center for Coherence and Quantum Optics, University of Rochester, Rochester, New York 14627, USA \\
\authormark{3} Racah Institute of Physics, The Hebrew University of Jerusalem, Jerusalem 91904, Israel 
}

\email{\authormark{*}daniel.lum@rochester.edu}




\begin{abstract}
We present an inexpensive architecture for converting a frequency-modulated continuous-wave LiDAR system into a compressive-sensing based depth-mapping camera. Instead of raster scanning to obtain depth-maps, compressive sensing is used to significantly reduce the number of measurements. 
Ideally, our approach requires two difference detectors. 
Due to the large flux entering the detectors, the signal amplification from heterodyne detection, and the effects of background subtraction from compressive sensing, the system can obtain higher signal-to-noise ratios over detector-array based schemes while scanning a scene faster than is possible through raster-scanning. 
Moreover, by efficiently storing only $2m$ data points from $m<n$ measurements of an $n$ pixel scene, we can easily extract depths by solving only two linear equations with efficient convex-optimization methods.
\end{abstract}




\section{Introduction}

Depth-mapping, also known as range-imaging or 3D imaging, is defined as a non-contact system used to produce a 3D representation of an object or scene \cite{cheok20103d}. 
Depth-maps are typically constructed through either triangulation and projection methods, which rely on the angle between an illumination source and a camera offset to infer distance, or on time-of-flight (TOF) LiDAR (light detection and ranging) techniques. LiDAR systems are combined with imaging systems to produce 2D maps or images where the pixel values correspond to depths.
 
Triangulation and projection methods, such as with Microsoft's Kinect \cite{smisek20133d}, are inexpensive and can easily operate at video frame rates but are often confined to ranges of a few meters with sub-millimeter uncertainty. Unfortunately, the uncertainty increases to centimeters after roughtly ten meters \cite{cheok20103d,khoshelham2012accuracy}. Phase-based triangulation systems also suffer from phase ambiguities \cite{geng2011structured}. Alternatively, LiDAR-based methods can obtain better depth-precision with less uncertainty, to within micrometers \cite{cheok20103d,behroozpour2017electronic}, or operate up to kilometer ranges. Thus, LiDAR based systems are chosen for active remote sensing applications when ranging accuracy or long-range sensing is desired. Additionally, technological advances in detectors, lasers, and micro-electro-mechanical devices are making it easier to incorporate depth-mapping into multiple technologies such as automated guided vehicles \cite{frank1993laser}, planetary exploration and docking \cite{amzajerdian2011lidar,stettner2010compact,4526302}, and unmanned aerial vehicle surveillance and mapping \cite{remondino2011uav}. Unfortunately, the need for higher resolution accurate depth-maps can be prohibitively expensive when designed with detector arrays or prohibitively slow when relying on raster scanning. 

Here we show how a relatively inexpensive, yet robust, architecture can convert one of the most accurate LiDAR systems available \cite{amann2001laser} into an efficient high-resolution depth-mapping system. Specifically, we present an architecture that transforms a frequency-modulated continuous-wave LiDAR into a depth-mapping device that uses compressive sensing (CS) to obtain high-resolution images. CS is a measurement technique that compressively acquires information significantly faster than raster-based scanning systems, trading time limitations for computational resources \cite{donoho2006compressed,duarte2008single,eldar2012compressed}. With the speed of modern-day computers, the compressive measurement trade-off results in a system that is easily scalable to higher resolutions, requires fewer measurements than raster-based techniques, obtains higher signal-to-noise ratios than systems using detector arrays, and is potentially less expensive than other systems having the same depth accuracy, precision, and image resolution. 

CS is already used in both pulsed and amplitude-modulated continuous-wave LiDAR systems (discussed in the next section). CS was combined with a pulsed time-of-flight (TOF) LiDAR using one single-photon-counting avalanche diode (SPAD), also known as a Geiger-mode APD, to yield $64\times 64$ pixel depth-images of objects concealed behind a burlap camouflage \cite{howland2011photon,colacco2012compressive}. CS was again combined with an SPAD-based TOF-LiDAR system to obtain $256\times 256$  pixel depth-maps, while also demonstrating background subtraction and $32\times 32$ pixel resolution compressive video \cite{howland2013photon}. Replacing the SPAD with a photodiode within a single-pixel pulsed LiDAR CS camera, as in \cite{sun2016single}, allowed for faster signal acquisition to obtain $128\times 128$ pixel resolution depth-maps and real-time video at $64\times 64$ pixel resolution. A similar method was executed with a coded aperture \cite{kadambi2015coded}. 
In addition to pulsed-based 3D compressive LiDAR imaging, continuous-wave amplitude-modulated LiDAR CS imaging \cite{Kirmani:11,conde2017compressive,antholzer2017framework}, ghost-imaging LiDAR \cite{doi:10.1063/1.4757874}, 3D reflectivity imaging \cite{yu2015three}, have been studied with similar results. The architecture we presented here, to our knowledge, is the first time CS has been applied to a frequency-modulated continuous-wave LiDAR system. 

Furthermore, we show how to efficiently store data from our compressive measurements and use a single total-variation minimization followed by two least-squares minimizations to efficiently reconstruct high-resolution depth-maps. We test our claims through computer simulations using debilitating levels of injected noise while still recovering depth maps. The accuracies of our reconstructions are then evaluated at varying levels of noise. Results show that the proposed method is robust against noise and laser-linewidth uncertainty.

\section{Current LiDAR and Depth-Mapping Technologies}

There are many types of LiDAR systems used in depth-mapping, yet there are two general categories for obtaining TOF information: pulsed LiDAR and continuous wave (CW) LiDAR \cite{horaud2016overview,remondino2013tof} (also known as discrete and full-waveform LiDAR, respectively \cite{lim2003lidar,mallet2009full}). 

Pulsed LiDAR systems use fast electronics and waveform generators to emit and detect pulses of light reflected from targets. Timing electronics measure the pulse's TOF directly to obtain distance. Depth resolution is dependent on the pulse width and timing resolution -- with more expensive systems obtaining sub-millimeter precision. Additionally, the pulse irradiance power is significantly higher than the background which allows the system to operate outside and over long range. Unfortunately, the high-pulse power can be dangerous and necessitates operating at an eye-insensitive frequency, such as the far-infrared, which adds additional expense. 
Generally, APDs or SPADs detect returning low-flux radiation and require raster scanning to form a depth-map. More expensive SPAD-arrays can quickly acquire depth-maps at the expense of SNR, but are currently limited to resolutions of $512\times 128$ \cite{Burri:14}.

Alternatively, CW-LiDAR systems continuously emit a signal to a target while keeping a reference signal, also known as a local oscillator. Because targets are continuously illuminated, they can operate with less power compared to the high peak-power of pulsed systems. CW-LiDAR systems either modulate the amplitude while keeping the frequency constant, as in amplitude-modulated continuous-wave (AMCW) LiDAR, or they modulate the frequency while keeping the amplitude constant, as in frequency-modulated continuous wave (FMCW) LiDAR. 

AMCW-LiDAR systems typically require high-speed radio frequency (RF) electronics to modulate the laser's intensity. However, low-speed electronics can be used to measure the return-signal after it has been demodulated. Examples of AMCW-LiDAR systems include phase-shift LiDAR, where the laser intensity is sinusoidally \cite{amann2001laser} or randomly modulated \cite{Takeuchi:86}. The TOF is obtained by convolving the demodulated local oscillator with the demodulated time-delayed return signal. 
Another popular AMCW system uses a linear-chirp of the laser's intensity and electronic heterodyne detection to generate a beat note proportional to the TOF \cite{stann1996intensity,batet2010intensity}. 

FMCW-LiDAR is the most precise LiDAR system available. The laser frequency is linearly chirped with a small fraction of the beam serving as a local oscillator. The modulation bandwidth is often larger than in linearly-chirped AMCW-LiDAR -- yielding superior depth resolution. Yet, high-speed electronics are not required for signal detection. Instead, detection requires an optical heterodyne measurement, using a slow square-law detector, to demodulate the return signal and generate a beat-note frequency that can be recorded by slower electronics. Unfortunately, FMCW-LiDAR is limited by the coherence length of the laser \cite{amann2001laser}. 

Both AMCW and FMCW linearly-chirped LiDAR systems can obtain similar or better depth precision compared to pulsed systems while operating with slower electronics. This is because the heterodyne beat note can be engineered to reside within the bandwidth of slow electronics and can be measured more precisely.
For this reason, there is much interest in using AMCW- and FMCW-LiDAR for depth-mapping. Our approach focuses on FMCW-LiDAR because of the the detection electronics are relatively easier to implement. 
Thus, a linearly-chirped FMCW-LiDAR that uses a high-bandwidth detector array for superior accuracy and a larger dynamic range is desirable. Yet, a high-bandwidth detector array may be prohibitively expensive and raster scanning a high-dimensional scene with a readily available high-bandwidth detector is oftentimes too slow. 
Our solution to this problem is to combine a compressive camera \cite{duarte2008single} with linearly-chirped FMCW-LiDAR system.

\section{FMCW-LiDAR Compressive Depth-Mapping}
\subsection{Single-Object FMCW-LiDAR}

Within an FMCW-LiDAR system, a laser is linearly swept from frequency $\nu_0$ to $\nu_f$ over a period $T$. The electric field $E$ takes the form
\begin{equation}
E(t) = A \exp\left(2\pi i\left[\nu_0 + \frac{\nu_f-\nu_0}{2T}t\right]t\right),
\label{eq:sweep}
\end{equation}
where $A$ is the field amplitude. It can be easily verified, by differentiation of Eq. (\ref{eq:sweep}), that the instantaneous laser frequency is $\nu_0+[(\nu_f-\nu_0)/T]t$. 

A small fraction of the laser's power is split off to form a local oscillator ($E_{\text{LO}}$) while the remainder is projected towards a target at an unknown distance $d$. The returning signal reflected off the target, with electric field $E_{\text{Sig}}(t-\tau)$ and time delay $\tau$, is combined with $E_{\text{LO}}$ and superimposed on a square-law detector. The resulting signal, as seen by an oscilloscope ($P_{\text{Scope}}$) and neglecting detector efficiency, is
\begin{equation}
P_{\text{Scope}}(t) = \frac{\epsilon_0 c}{2}\left[A_{\text{Sig}}^2 + A_{\text{LO}}^2 +
 2A_{\text{LO}}A_{\text{Sig}}\sin\left(2\pi\frac{\Delta\nu\tau}{T}t+\phi\right)\right],
\label{eq:singleLIDAR} 
\end{equation}
where $\phi = \nu_0\tau - \frac{\Delta\nu}{2T}\tau^2$ is a constant phase, $\epsilon_0$ is the permittivity of free space, $c$ is the speed of light in vacuum, and $\Delta\nu = \nu_f-\nu_0$. The alternating-current (AC) component within Eq. (\ref{eq:singleLIDAR}) oscillates at the beat-note frequency $\nu = \Delta\nu\tau/T$ from which a distance-to-target can be calculated from
\begin{equation}
d = \frac{\nu T c}{2\Delta\nu}.
\label{eq:dist}
\end{equation}

\subsection{Multiple-Object FMCW-LiDAR}

The detection scheme should be altered when using a broad illumination profile to detect multiple targets simultaneously. Just as beat notes exist between the local oscillator and signal, there now exist beat notes between reflected signals at varying depths that will only inject noise into the final readout. Balanced heterodyne detection is used to overcome this additional noise source. A 50/50 beam-splitter is used to first mix $E_{\text{LO}}$ and $E_{\text{Sig}}$. Both beamsplitter outputs, containing equal powers, are then individually detected and differenced with a difference-detector. 

Mathematically, $E_{\mathrm{LO}}$ is the same as before, but the return signal from $j$ objects at different depths takes the form
\begin{equation}
\sum\limits_{j} E_{\text{Sig}}(t-\tau_j) = \sum\limits_{j} A_{j}\exp\left(2\pi i\left[\nu_0+\frac{\Delta\nu}{2T}\left(t-\tau_j\right)\right]\left(t-\tau_j\right)\right),
\end{equation} 
where we have labeled each object with an identifier index $j$ that introduces a time delay $\tau_j$ and amplitude $A_j$, depending on the reflectivity of the object. After mixing at the beamsplitter, the optical power from the difference detector, as seen by an oscilloscope ($P_{\text{Scope}}$), again neglecting detector efficiency, is  
\begin{align}
P_{\text{Scope}}(t)&=\frac{\epsilon_0 c}{2}\left(\left|\frac{E_{\text{LO}}\left(t\right) + i\sum_{j} E_{\text{Sig}}\left(t-\tau_j\right)}{\sqrt{2}}\right|^2 - \left|\frac{iE_{\text{LO}}\left(t\right) + 
\sum_{j} E_{\text{Sig}}\left(t-\tau_j\right)}{\sqrt{2}}\right|^2\right) \\
&= \epsilon_0 c\sum\limits_{j} A_{\text{LO}}A_{j}\sin\left(2\pi\frac{\Delta\nu\tau_j}{T}t + \phi_j\right),
\label{eq:heterodyneSig}
\end{align}
where $\phi_j = \nu_0\tau_j - \frac{\Delta\nu}{2T}\tau_j^2$ is a constant phase. The sum-frequency signal components are filtered by the slow-bandwidth detectors. The resulting balanced heterodyne signal in Eq. (\ref{eq:heterodyneSig}) is still amplified by the local oscillator and contains AC signals at only the frequencies $\nu_j = \Delta\nu\tau_j/T$. Again, these frequencies can be converted to distance using Eq. (\ref{eq:dist}).

\subsection{Compressive imaging background}

Compressive imaging is a technique that trades a measurement problem for a computational reconstruction within limited-resource systems. Perhaps, the most well known example of a limited-resource system is the Rice single-pixel camera \cite{duarte2008single}. Instead of raster scanning a single-pixel to form an $n$-pixel resolution image $\vect{x}$, i.e. $\vect{x}\in\mathbb{R}^{n}$, an $n$-pixel digital micro-mirror device (DMD) takes $m\ll n$ random projections of the image. The set of all DMD patterns can be arranged into a sensing matrix $\vect{A}\in\mathbb{R}^{m\times n}$, and the measurement is modeled as a linear operation to form a measurement vector $\vect{y}\in\mathbb{R}^{m}$ such that $\vect{y}=\vect{Ax}$.

Once $\vect{y}$ has been obtained, we must reconstruct $\vect{x}$ within an undersampled system. As there are an infinite number of viable solutions for $\vect{x}$ within the problem $\vect{y} = \vect{Ax}$, CS requires additional information about the signal according to a previously-known function $g(\vect{x})$. The function $g(\vect{x})$ first transforms $\vect{x}$ into a sparse or approximately sparse representation, i.e. a representation with few, or approximately few, non-zero components. The function $g(\vect{x})$ then uses the $L^1$-norm to return a scalar.

Within image-reconstruction problems, $g(\vect{x})$ is often the total-variation (TV) operator because it provides a sparse representation by finding an image's edges. Anisotropic TV is defined as $\text{TV}(\vect{x}) = \parallel [\nabla_x^T, \nabla_y^T]^T \vect{x}\parallel_1$, where $\nabla$ is a finite difference operator that acts on an image's Cartesian elements ($x$ and $y$) and $\ast^T$ is the transpose of $\ast$. Note that the $L^p$-norm of $\vect{x}$ is defined as $\parallel \vect{x} \parallel_p = (\sum_i |\vect{x}_i|^p )^{1/p}$. Thus, we solve the following TV-minimization problem:
\begin{equation}
\text{arg}\,\min\limits_{\hat{\vect{x}}\in\mathbb{R}^n}\parallel \vect{A}\hat{\vect{x}} - \vect{y} \parallel_2^2 + \alpha \mathrm{TV}\left(\hat{\vect{x}}\right),
\label{eq:TV}
\end{equation}
where the first term is a least-squares fitting parameter consistent with the measurements $\vect{y}$, the second term is a sparsity promoting parameter, and $\alpha$ is a weighting constant.

\subsection{Compressive FMCW-LiDAR depth-mapping theory} \label{ssec:ReconTheory}

\begin{figure}
\centering\includegraphics[width=.8\textwidth]{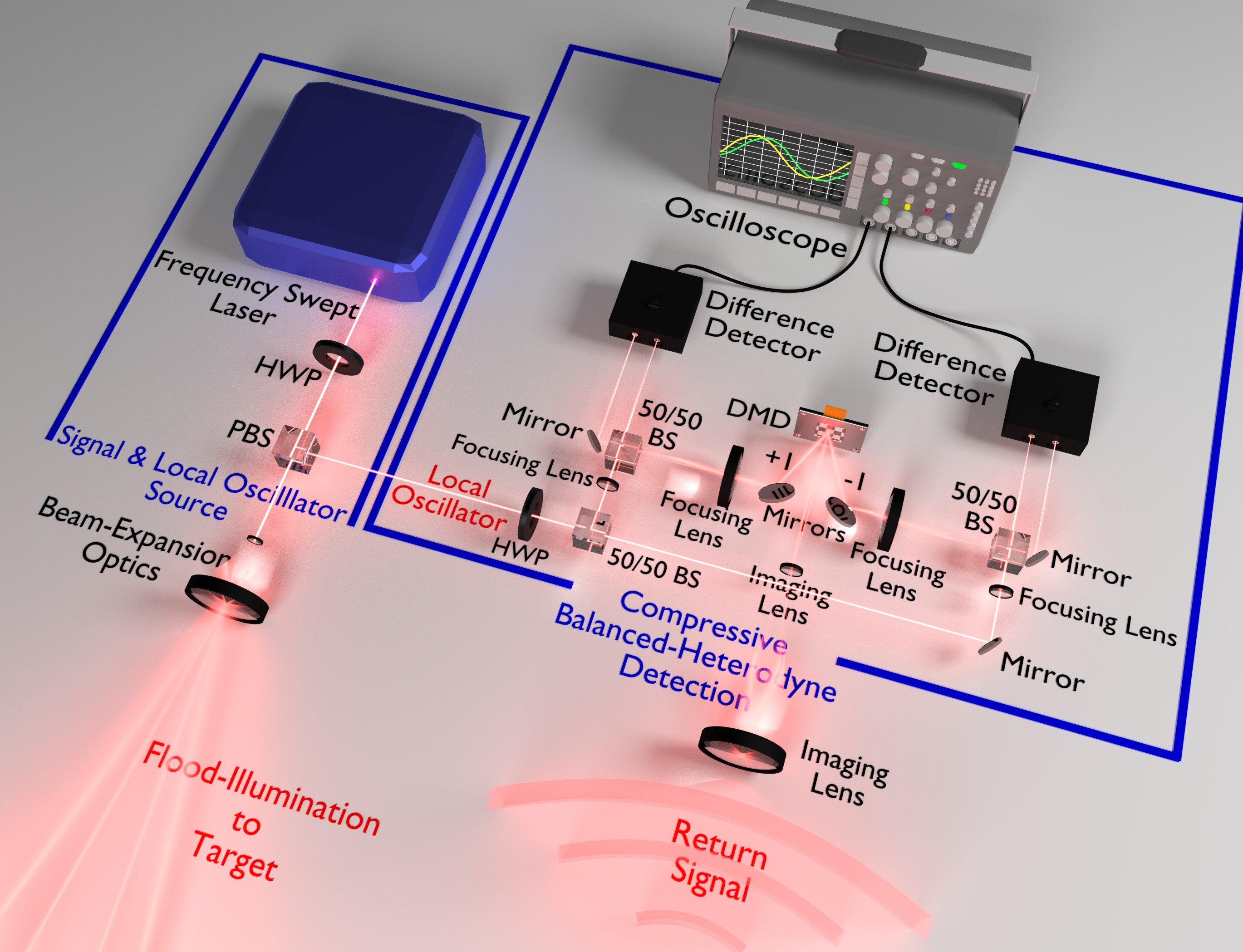}
\caption{(Proposed Experiment) HWP: half-wave plate, PBS: polarizing beam-splitter, DMD: digital micro-mirror device, BS: beam-splitter. A linearly chirped laser is split into two beams, designated as a local-oscillator and a signal, via a HWP and a PBS. The signal illuminates a target scene and the reflected radiation is used to image the scene onto a DMD. The DMD takes pseudo-random spatial projections, consisting of $\pm 1$ pixel values, and directs the projections to balanced-heterodyne detectors using the local-oscillator.}
\label{fig:setup}
\end{figure}

The proposed experimental diagram is shown in Fig. \ref{fig:setup}. A balanced-heterodyne-detection based FMCW-LiDAR system for multiple-object ranging is combined with a DMD-based compressive-imaging system to obtain transverse spatial information more efficiently than raster scanning. A linear-frequency chirped laser broadly illuminates a scene and the reflected radiation is imaged onto a DMD. 
The DMD can take projections using $\pm 1$ pixel values, meaning the sensing matrix $\vect{A}$ is limited to $\pm 1$ values. Because the 1 and -1 DMD pixel values reflect light in different directions, we split the acquisition process into separate $(1,0)$ and $(0,1)$ balanced heterodyne detections. A readout instrument, such as an oscilloscope, records the two signals. The time-dependent readout signals are then Fourier transformed -- keeping only the positive frequency components and the amplitudes are differenced to arrive at a $(1,-1)$ balanced heterodyne projection frequency representation. 
The measurement vector construction and depth-map reconstruction processes are detailed in this section.

Let a three-dimensional scene $\vect{x}$ of depth $L$ be discretized into $N$ depths and $n$ transverse spatial pixels containing objects at $j$ depths. Let a 2D image of the scene $\vect{x}$ at depth $j$ be represented by a real one-dimensional vector $\vect{x}_{Ij} \in \mathbb{R}^{n}$. 
The $l^{\mathrm{th}}$ pixel in $\vect{x}_{Ij}$ is represented as $\vect{x}_{Ij}^{l}$.
The scene at depth $j$ (i.e. $\vect{x}_{Ij}$) yields a time-dependent heterodyne signal of $\sum_{l=1}^n \vect{x}_{Ij}^l = \epsilon_0cA_{\mathrm{LO}}A_{j}\sin\left(2\pi\nu_jt+\phi_j\right)$, where $\nu_j=\Delta\nu\tau_j/T$. Let us also assume that we are time-sampling at the Nyquist rate needed to see an object at a maximum depth $L$ having a beat-note frequency of $2\Delta\nu L / (Tc)$, meaning we acquire $2N$ data points when sampling at a frequency of $4\Delta\nu L / (Tc)$ over a period $T$. This is the minimum sampling requirement. Within an experiment, sampling faster than the Nyquist rate will enable the accurate recovery of both the frequency and amplitude at depth $L$.
Let a single pattern to be projected onto the imaged scene be chosen from the sensing matrix $\vect{A}\in\mathbb{R}^{m\times n}$ and be represented by a real one-dimensional vector $\vect{A}_k \in \mathbb{R}^n$ for $k = 1,2,...m$, where $m < n$. For efficent compression, we require $m\ll n$. When including the $2N$ time samples, the heterodyne signal is now represented as a matrix of $n$ pixels by $2N$ time points, $\vect{P}_{\mathrm{Scope}}(t)\in\mathbb{R}^{m\times 2N}$, such that
\begin{equation}
\vect{P}_{\mathrm{Scope}}(t) = \epsilon_0 c\sum\limits_{j} A_{\text{LO}}\vect{A}\vect{x}_{Ij}\sin\left(2\pi\frac{\Delta\nu\tau_j}{T}t + \phi_j\right).
\end{equation}

To obtain TOF information, it is helpful to work in the Fourier domain. The absolute value of the first $N$ elements in the Fourier transform of $\vect{P}_{\mathrm{Scope}}(t)$ returns the compressive measurement matrix of positive frequency amplitudes, $\vect{y}_{I}(\nu_+)=\left\lvert\mathscr{F}\left[\vect{P}_{\mathrm{Scope}}(t)\right]_{+}\right\rvert \in\mathbb{R}^{m\times N}$, where $\mathscr{F}\left[\ast\right]_{+}$ returns the positive-frequency components of $\ast$. Thus, $\vect{y}_I(\nu_+)$ consists of $N$ different measurement vectors of length $m$. Each of the $N$ measurement vectors contain compressed transverse information for a particular depth.

While individual images at a particular depth can be reconstructed from $\vect{y}_I(\nu_+)$, it is inefficient based on the size of the measurement matrix and the number of needed reconstructions. A more practical method with a significantly smaller memory footprint and far fewer reconstructions requires that we trace out the frequency dependence and generate only two measurement vectors of length $m$, designated as $\vect{y}_I$ and $\vect{y}_{I\nu}\in\mathbb{R}^m$ such that
\begin{align}
\vect{y}_I &= \sum\limits_{\nu_+ = 1}^N \left\lvert\mathscr{F}\left[\vect{P}_{\mathrm{Scope}}(t)\right]_{+}\right\rvert \label{eq:yi}\\
\vect{y}_{I\nu} &= \sum\limits_{\nu_+ = 1}^N \left\lvert\mathscr{F}\left[\vect{P}_{\mathrm{Scope}}(t)\right]_{+}\right\rvert \vect{\nu_+}, \label{eq:yvi}
\end{align}
where $\vect{\nu_+}\in\mathbb{R}^{N\times N}$ is a diagonal matrix of frequency indices that weights each frequency amplitude of $\vect{y}_I(\nu_+)$ by its own frequency ($\nu$) before summation.

To summarize, the measurement vector $\vect{y}_I$ is formed by summing the positive-frequency amplitudes, per projection, and is equivalent to a typical compressive-imaging measurement. The measurement vector $\vect{y}_{I\nu}$ requires that each positive-frequency amplitude first be weighted by its frequency before summation. A depth map $\vect{d}$ can be extracted by reconstructing the transverse image profiles ($\hat{\vect{x}}_I, \hat{\vect{x}}_{I\nu}\in\mathbb{R}^n$) and then performing a scaled element-wise division as 
\begin{equation}
\vect{d} = \frac{\hat{\vect{x}}_{I\nu}}{\hat{\vect{x}}_I}\frac{T c}{2 \Delta\nu},
\label{eq:depth}
\end{equation}
where $\hat{\vect{x}}_{I\nu}/\hat{\vect{x}}_I$ is an element-wise division for elements $>0$. A similar technique designed for photon-counting with time-tagging was implemented in \cite{howland2011photon}.

\subsection{Compressive gains}
\label{ssec:gains}

Now that the projective aspect of the CS measurement is understood, it is beneficial to emphasize the advantages compressive depth-mapping offers over raster and detector-array based systems. In short, projective measurements result in eye-safety and SNR enhancements.  

Regarding eye-safety, it is easier to render the illuminating radiation of CS methods eye-safe when compared to raster-scanning methods. As demonstrated in \cite{sun2016single}, a broad illumination-profile can have a lower intensity across the eye when compared against raster-scanning with a tightly focused laser. 

The SNR is also enhanced in compressive methods, particularly in comparison to detector arrays, and has been extensively studied in \cite{yu2016compressive,soldevila2016computational,cevher2008compressive,yu2014complementary}. In the case of depth-mapping with broad-illumination profiles, lower return flux will be an issue. If the flux rate $R$ of the returning radiation is divided among $\alpha$ pixels within a $\beta$-pixel detector array (assuming $\alpha\leq \beta$ reflective pixels within the scene are imaged onto the $\beta$-pixel detector array), the number of photons received by a single detector after an integration time $t$ will be approximately $Rt/\alpha$. In the single-detector compressive case, roughly half of the photons will make it to the detector due to the random projection. If we want the compressive method -- requiring $m$ different projective measurements -- to have the same total integration time $t$, then the integration time for a single projection must be $t/m$. Assuming a DMD can operate at this rate, the number of photons received by the detector during a single compressive projection will be $Rt/(2m)$. Because compression allows for $2m < \alpha \leq \beta$, the increased flux would reduce shot-noise uncertainty. 

In addition to a higher SNR from using fewer pixels, the system presented here benefits from both balanced heterodyne detection and background subtraction. Heterodyne detection will amplify the return radiation with the local oscillator. Background subtraction from the $(1,-1)$ projections will cancel correlated noise \cite{yu2016compressive,cevher2008compressive,yu2014complementary,Gerrits:17}. Here, we define correlated noise as identical noise sources present in the $(+1,0)$ and $(0,-1)$ projections from the DMD. Compressive sensing with $(1,-1)$ valued patterns has been shown to successfully recover signals even when the background is larger than the signal itself. This can be easily seen by adding complex valued noise terms to the signal and then propagating them through the system up to the generation of the measurement vectors. Perfectly correlated noise sources, such as uniform background radiation, will cancel. Partially correlated noise sources, such as spatially varying background radiation, will only partially cancel. The computer simulations presented below inject uncorrelated noise and must be removed through denoising algorithms.     


\section{Computer Simulation}

\subsection{Simulation parameters} \label{ssec:param}

\begin{figure}
\centering\includegraphics[width=.66\textwidth]{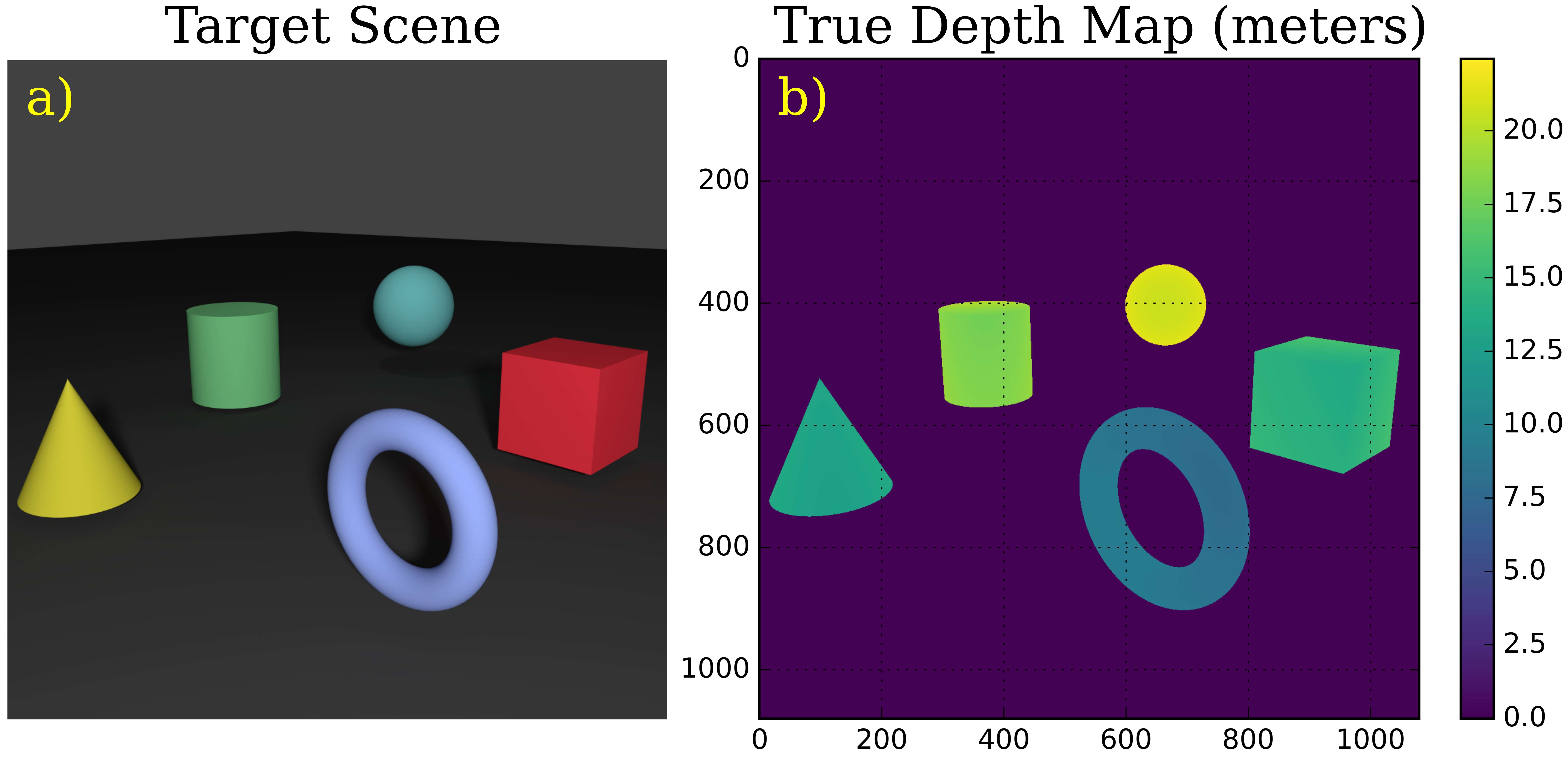}
\caption{Image (a) presents a 3-dimensional scene composed of Lambertian-scattering targets. Image (b) presents the depth map we wish to compressively recover.}
\label{fig:original}
\end{figure}

\begin{figure}
\centering\includegraphics[width=.78\textwidth]{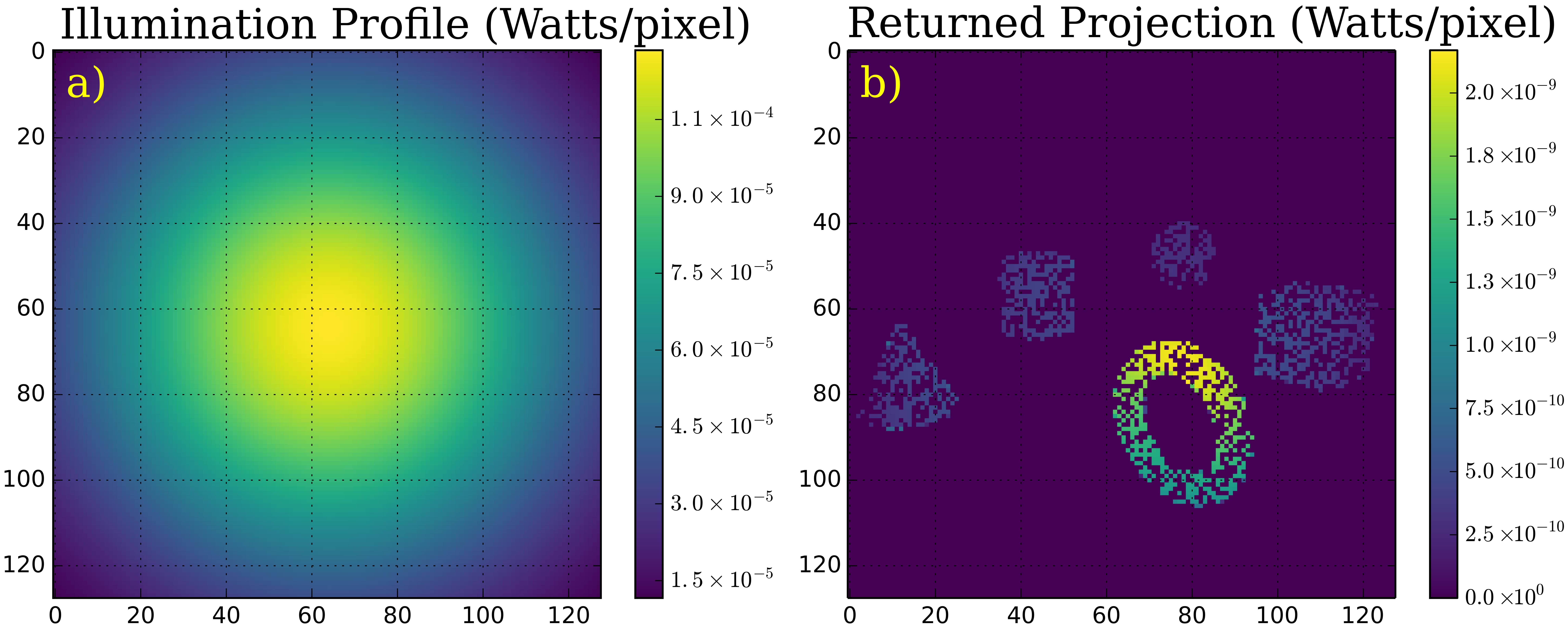}
\caption{Image (a) presents an illumination profile that consists of 1 Watt in total power. The scene is discretized into a $128 \times 128$ pixel resolution scene with a maximum power per pixel of 119 $\mu$W. When using a 2 inch collection optic and modeling each object as a Lambertian scatter, image (b) presents a single projection taken by the DMD of the reflected radiation as seen by one detector. The maximum power per pixel is of order nanowatts.}
\label{fig:illumination}
\end{figure}

A noisy simulation of compressive depth-mapping is presented within this section. The simulation presents detailed steps needed to acquire depth-maps from signals disrupted by both laser-linewidth uncertainty and Gaussian white noise when compressively measuring the scene presented in Fig. \ref{fig:original}. We assume the scene contains Lambertian-scattering objects and is illuminated with 1 W of 780 nm light using an illumination profile shown in Fig. \ref{fig:illumination}. With a large enough angular spread of optical power, the system can be rendered eye-safe at an appropriate distance from the source. We discretize the scene with a $128\times 128$ pixel resolution DMD ($n = 16384$) operating at 1 kHz and use a 2 inch collection optic according to the diagram in Fig. \ref{fig:setup}. Figure \ref{fig:illumination} presents an example DMD projection of the returned radiation intensities before the balanced heterodyne detection as seen by only one of the detectors (corresponding to a (1,0) projection). A local oscillator of 100 $\mu$W is mixed with the returned image for the balanced heterodyne detection. 

To model the frequency sweep, we use experimental parameters taken from \cite{cheok20103d}  and \cite{Satyan:09} which state that a 100 GHz linear sweep over 1 ms is realizable for various laser sources. To bias the simulation towards worse performance, we model the laser linewidth with a 1 MHz Lorentzian full-width half-maximum (FWHM), meaning the beat note on our detectors will have a 2 MHz FWHM Lorentzian linewidth \cite{nazarathy1989spectral}. A narrower linewidth can be obtained with many lasers on the market today, but we bias the simulation towards worse operating parameters to show the architecture's robustness. The simulated laser has a coherence length of $c/(\pi \Delta\nu_\mathrm{FWHM}) \approx 95$ m, where $\Delta\nu_{\mathrm{FWHM}} = 1$ MHz.

FMCW-LiDAR can use relatively inexpensive measurement electronics while still providing decent resolution at short to medium range. Using a 100 GHz frequency sweep over 1 ms leads to a beat note of 16.67 MHz from an object 25 m away. Keeping with the idea of slower electronics, we model the detection scheme using balanced-detectors and an oscilloscope that can see 16.65 MHz. The simulated oscilloscope sampled at 33.3 MHz over 1 ms, corresponding to a record-length of $33.3\times 10^3$ samples per frequency sweep. The frequency resolution is 1 kHz with a maximum depth resolution of 1.5 mm. The furthest object in our test scene is located at approximately 22 m. 

We also assume a 1 ms integration time per DMD projection, corresponding to one period of the frequency sweep per projection. We model the linear frequency chirp using a sawtooth function for simplicity, as opposed to more commonly used triangle function in experiments. 

The sensing matrix used for the compressive sampling is based on a randomized Sylvester-Hadamard matrix displays values of $\pm 1$. We use two detectors to acquire the information with one projection -- resulting in only $m$ projections for an $m$-row sampling matrix. Alternatively, a cheaper single-detector version can also be implemented, but it must display two patterns for every row of the sensing matrix -- resulting in $2m$ projections.  While the alternate detection scheme requires twice as many measurements, we still operate in the regime where $2m \ll n$.

\subsection{Compressive measurements} \label{ssec:measure}

\begin{figure}
\centering\includegraphics[width=.9\textwidth]{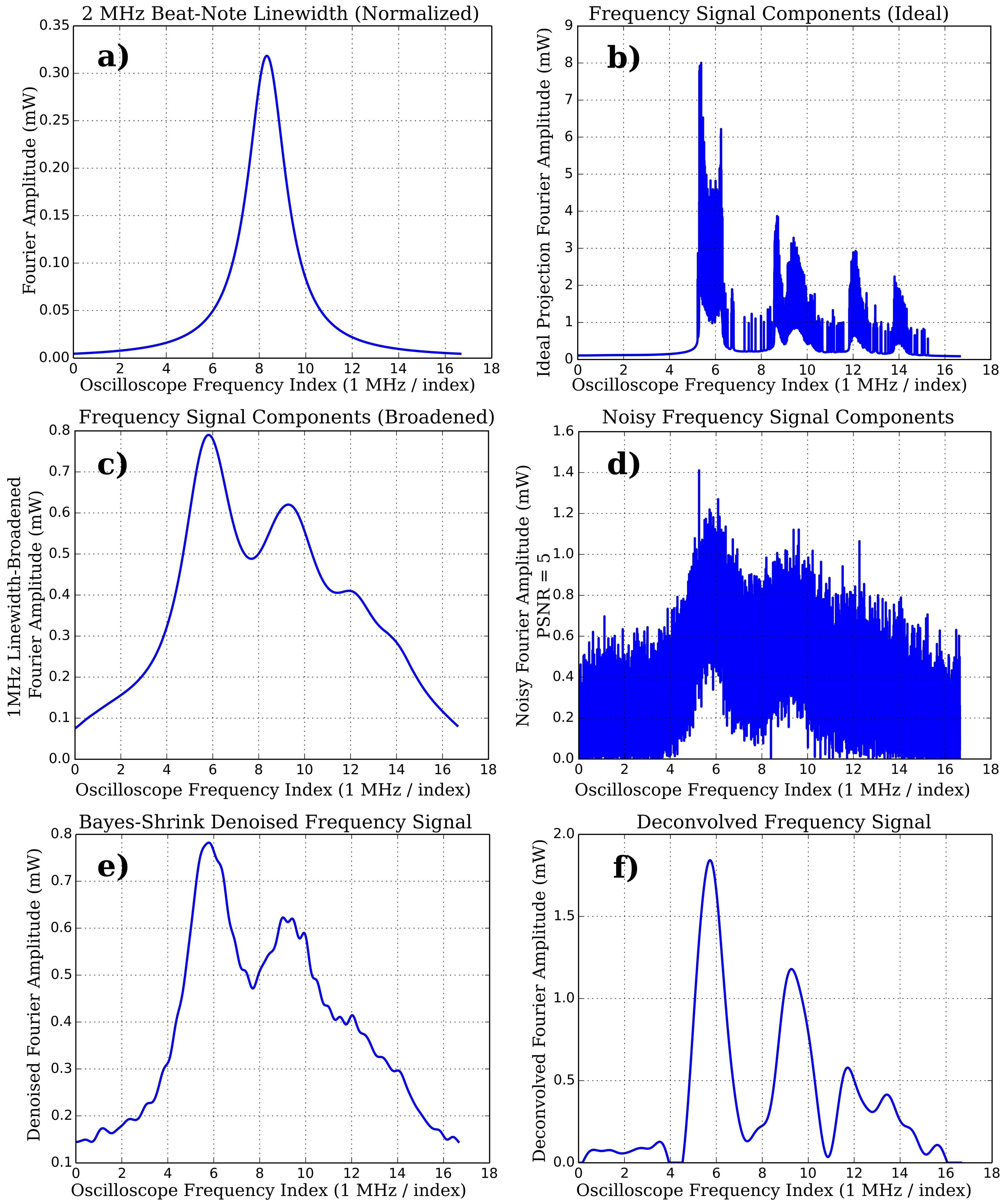}
\caption{Image (a) shows the 2 MHz Lorentzian linewidth on the beat notes expected from a laser's 1 MHz Lorentzian linewidth as seen by an oscilloscope sampling at 33.3 MHz. Image (b) shows the noiseless, positive-frequency components from a single noiseless projection from Fig. \ref{fig:illumination} when using a zero-linewidth laser. When accounting for a 2 MHz beat-note linewidth, image (c) shows the broadened expected frequency components. When including a peak SNR = 5  (based on the SNR of the brightest object or pixel), image (d) presents a realistic noisy signal one would measure in an experiment. Image (e) presents the result of denoising the signal in (d) with a Bayes-shrink denoising filter using a symlet-20 wavelet decomposition. Image (f) presents the result of deconvolving the 2 MHz Lorentzian linewidth with the denoised signal in (e) using a Weiner filter. Image (f) is the cleaned signal from the (1,0) projection. A similarly cleaned result from the (0,1) signal will then be subtracted from (f) and then used to form $\vect{y}_I$ and $\vect{y}_{I\nu}$. }
\label{fig:signal}
\end{figure}

To simulate data acquisition in a noisy environment, we add Gaussian white noise in addition to laser-linewidth uncertainty. This section explains how the Fourier-transformed oscilloscope amplitudes are processed before applying the operations found in Eqs. (\ref{eq:yi}) and (\ref{eq:yvi}). 

Figure \ref{fig:signal} explains the simulated noise addition and removal process. Image (a) depicts a 2 MHz normalized beat note Lorentzian linewidth uncertainty in terms frequency components. Image (b) presents a single ideal DMD projection, using a 1 ms integration time, that has not been distorted by any form of noise. Image (b) clearly depicts 5 main peaks, one per object. However, beat-note linewidth uncertainty will blur these features together such that image (c) contains frequency uncertainty. In addition to laser-linewidth, varying levels of Gaussian white noise are injected into image (c). Image (d) presents a typical noisy signal one would expect to see for a peak signal-to-noise ratio (PSNR) of 5, meaning the SNR of the brightest Lorentzian-broadened Fourier component has an SNR equal to 5. SNR is defined as the signal's boradened amplitude divided by the standard deviation of Gaussian white noise. Note that the farthest object has an SNR at or below unity.

To clean a typical signal seen by only one detector, as in image (d), we use a BayesShrink filter \cite{BayesShrink} with a symlet-20 wavelet decomposition (see the PyWavelets library for more details). BayesShrink is a filter designed to suppress Gaussian noise using a level-dependent soft-threshold at each level of a wavelet decomposition. Because a symlet-20 wavelet is reasonably close in shape to the original signal components found in image (b), it has excellent denoising properties for these particular data sets. Alternate wavelets or denoising algorithms can be implement and may demonstrate better performance. Image (e) is the result of applying a symlet-20 based BayesShrink filter. A Weiner deconvolution filter \cite{wiener1949extrapolation} is then used to deconvolve the known Lorentzian linewidth of image (a) with the denoised signal in (e) to produce signal in (f). Image (f) constitutes only one of the needed projections, i.e. (1,0). The process must also be applied to the signal from the second detector consisting of (0,1) projection. The two cleaned signals are then subtracted to produce a (1,-1) projection. The differenced signal is then used to form a single element of each measurement vector $\vect{y}_{I}$ and $\vect{y}_{I\nu}$ using Eqs. \ref{eq:yi} and \ref{eq:yvi}. 


\subsection{Depth-map reconstruction} \label{ssec:reconstruct}

\begin{figure}
\centering\includegraphics[width=\textwidth]{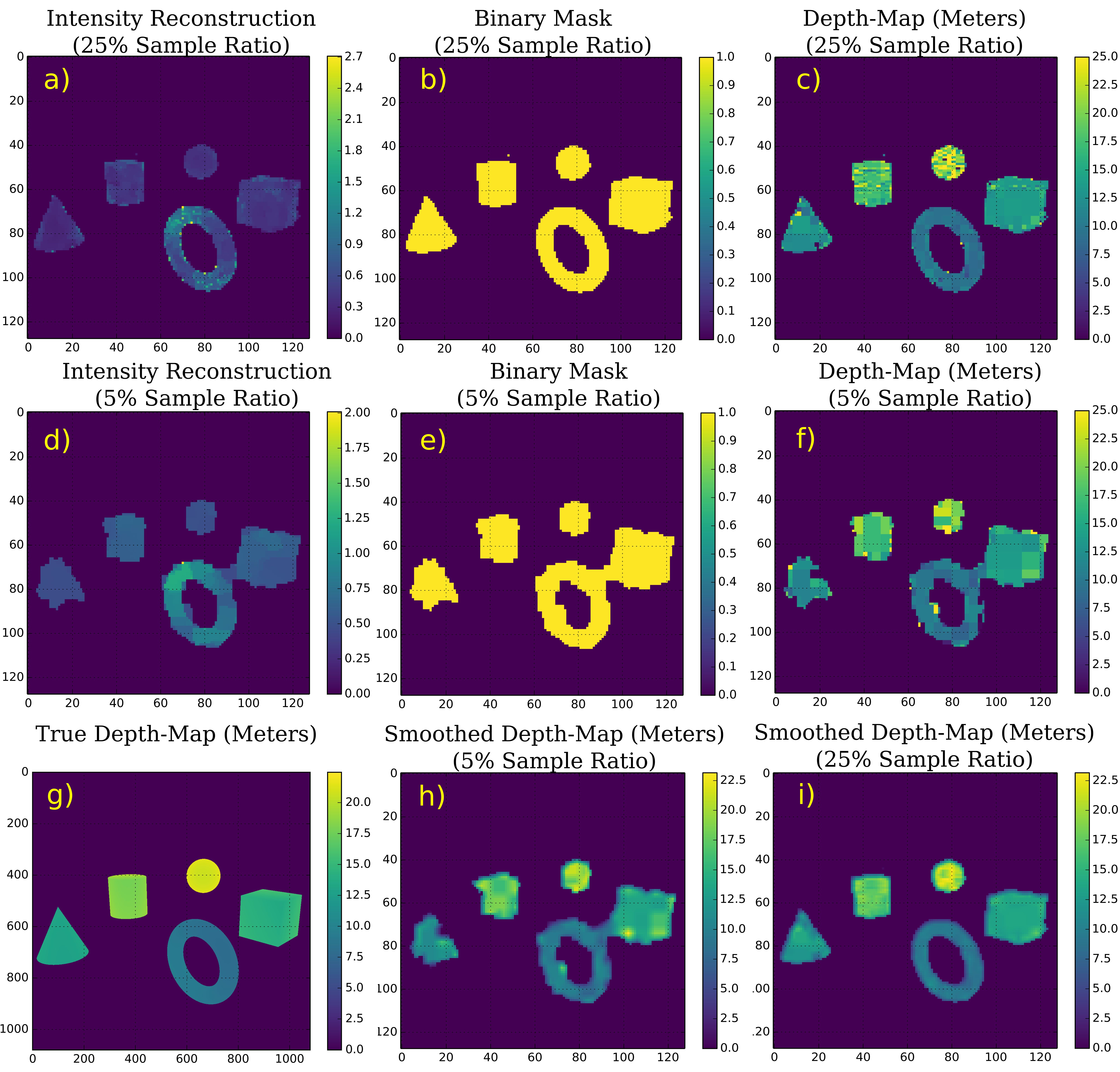}
\caption{When using a PSNR = 5, image (a) shows a typical total-variation minimization reconstruction from a $25\%$ sample-ratio intensity-measurement vector $\vect{y}_I$. Shapes are identified, but the pixel values are incorrect. Image (b) is a binary mask generated by hard-thresholding image (a). After representing image (b) in a sparse basis, such as a Haar-wavelet decomposition, least-squares can be performed on the $m/3$ largest signal components to construct $\vect{x}_I$ and $\vect{x}_{I\nu}$. After applying Eq. (\ref{eq:depth}), a depth map is presented in image (c). Images (d), (e), and (f) demonstrate the same process for a $5\%$ sample ratio, again with PSNR = 5. Image (g) is the true depth-map presented for easy comparison. Images (h) and (i) are smoothed depth-maps after applying a $4\!\times\! 4$ pixel averaging kernel to depth-maps (f) and (c), respectively. }
\label{fig:results}
\end{figure}

Once the process outlined in section \ref{ssec:measure} has been repeated $m$ times, depth maps are reconstructed from $\vect{y}_{I}$ and $\vect{y}_{I\nu}$ using the procedure outlined in section \ref{ssec:ReconTheory}. Details to the recovery of $\hat{\vect{x}}_I$ and $\hat{\vect{x}}_{I\nu}$ are presented in this section.

Ideally, one would only solve Eq. (\ref{eq:TV}) twice to obtain $\hat{\vect{x}}_I$ and $\hat{\vect{x}}_{I\nu}$ and then apply Eq. (\ref{eq:depth}) to extract high-resolution depth-maps. Unfortunately, TV-minimization will not always return accurate pixel values, particularly at low sample percentages, and will result in depth-maps with well-defined objects but at incorrect depths. While TV-minimization excels at finding objects when making $\alpha>1$, the least-squares term that ensure accurate pixel values suffers. To obtain accurate depth maps, we must locate objects accurately \emph{and} obtain accurate pixel values.

Instead, we only apply one TV-minimization to locate the objects within $\hat{\vect{x}}_{I}$. After hard thresholding the image returned from TV-minimization, if needed to eliminate background noise, we are left with an image of objects with incorrect pixel values. An example of this step can be found in images (a) and (d) of Fig. \ref{fig:results}, where image (a) was obtained from $m=.25n$ measurements and image (d) was obtained from $m=.05n$ measurements. From $\hat{\vect{x}}_{I}$, we generate a binary mask $\vect{M}$, as demonstrated in images (b) and (e) that tells us the locations of interest.

Once we have a binary mask $\vect{M}$, a unitary transform $\Psi$ converts $\vect{M}$ into a sparse representation $\vect{s}$ such that $\vect{s} = \Psi\vect{M}$. In our simulations, $\Psi$ is a wavelet transformation using a Haar wavelet. We tested other wavelets with similar results. We then keep only $m/3$ of the largest signal components of $\vect{s}$ to form a vector $\vect{s}_{m/3} = \vect{P}\vect{s}$, where the operator $\vect{P}\in\mathbb{R}^{m/3\times n}$ is a sub-sampled identity matrix that extracts the $m/3$ largest signal components of $\vect{s}$. We chose $m/3$ rather than a larger percentage of $m$ based on empirical results. In essence, $\vect{s}_{m/3}$ is the support of $\vect{s}$, i.e. the significant nonzero components of $\vect{s}$ containing the transverse spatial information of our depth-map. Because we have a vector $\vect{s}_{m/3}\in\mathbb{R}^{m/3}$ and two measurement vectors $\vect{y}_{I}\in\mathbb{R}^m$ and $\vect{y}_{I\nu}\in\mathbb{R}^m$, we can perform least-squares on the now-overdetermined system. Specifically, we apply the operations
\begin{align}
\vect{s}_{m/3} &= \vect{P}\Psi\hat{\mathrm{M}}\left(\text{arg}\,\min\limits_{\vect{x}_{I}\in\mathbb{R}^n}\parallel \vect{A}\vect{x}_I - \vect{y}_I \parallel_2^2 + \alpha \mathrm{TV}\left(\vect{x}_I\right)\right) \label{eq:TVxI}\\
\hat{\vect{x}}_I &= \vect{M}\Psi^{-1}\vect{P}^T\left(\mathrm{arg}\,\min\limits_{\vect{s}_{m/3}}\parallel \vect{A}\Psi^{-1}\vect{P}^T\vect{s}_{m/3}-\vect{y}_I\parallel_2^2\right) \label{eq:LSxI} \\
\hat{\vect{x}}_{I\nu} &= \vect{M}\Psi^{-1}\vect{P}^T\left(\mathrm{arg}\,\min\limits_{\vect{s}_{m/3}}\parallel \vect{A}\Psi^{-1}\vect{P}^T\vect{s}_{m/3}-\vect{y}_{I\nu}\parallel_2^2\right) \label{eq:LSxInu},
\end{align}
where the operation $\hat{\mathrm{M}}$ returns a mask $\vect{M}$. Note that $\vect{P}^T$ is the transpose of $\vect{P}$ and $\Psi^{-1}$ is the inverse wavelet transform. Also notice that $\vect{s}_{m/3}$ and $\vect{P}$ contain the transverse spatial information of interest within our depth-map and are used in both Eqs. (\ref{eq:LSxI}) and (\ref{eq:LSxInu}). They allow us to obtain an accurate heterodyne amplitude image $\hat{\vect{x}}_I$ (from $\vect{y}_I$) and an accurate frequency weighted heterodyne amplitude image $\hat{\vect{x}}_{I\nu}$ (from $\vect{y}_{I\nu}$). 

The total variation minimization was performed with a custom augmented Lagrangian algorithm written in Python. It is based on the the work in \cite{yin2010practical} but is modified to use fast-Hadamard transforms. Additionally, the gradient operators within the TV operator use periodic boundary conditions that enables the entire algorithm to function with only fast-Hadamard transforms, fast-Fourier transforms, and soft-thresholding. The least-squares minimizations in Eqs. (\ref{eq:LSxI}) and (\ref{eq:LSxInu}) are performed by the Broyden-Fletcher-Goldfarb-Shanno (BFGS) algorithm \cite{fletcher2013practical,byrd1995limited} -- a fast quasi-Newton method. Depth maps are obtained by applying Eq. (\ref{eq:depth}). Example depth maps are found in images (c) and (f) of Fig. \ref{fig:results}. If desired, light smoothing can be applied to obtain images (h) and (i).

The reconstruction algorithms recovered the low-spatial-frequency components of the depth-maps in Fig. \ref{fig:results}, and arguably recovered a low-resolution depth-map. However, missing details can be retrieved when using a higher resolution DMD, sampling with higher sampling ratios, or applying adaptive techniques \cite{AdaptiveCS}.   

\subsection{Results}

\begin{figure}
\centering\includegraphics[width=\textwidth]{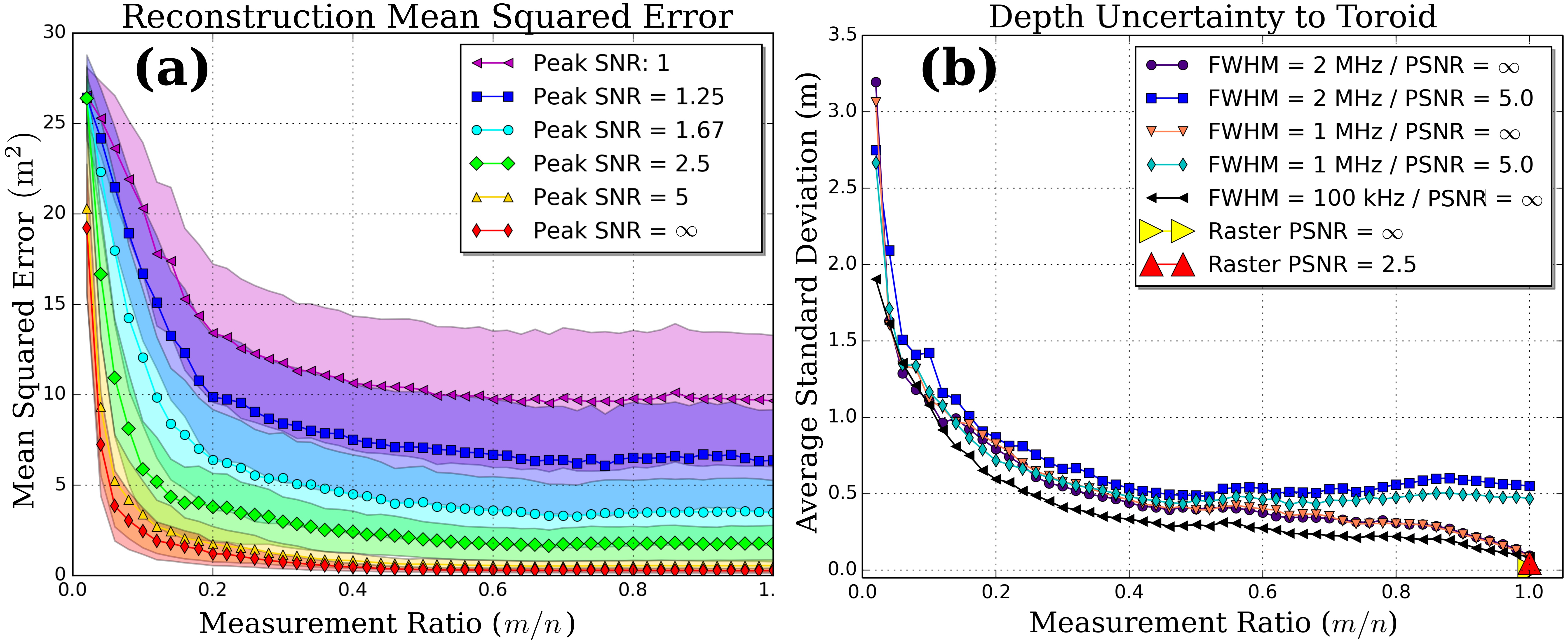}
\caption{(a) Reconstruction Mean-Squared Error: Each data point corresponds to the average mean squared error from 10 different reconstructions of different data sets for varying sample ratios ($m/n$) and peak-SNR cases. The standard error for each mean is given by the shaded regions. (b) Depth Uncertainty to Toroid: Each data point represents the average uncertainty in depth to the toroid within Fig. \ref{fig:original}. Points were calculated by first considering the standard deviation of each pixel over 10 different reconstructions. Standard deviations associated with only the toroid pixels were then averaged. Different peak-SNR scenarios, FWHM Lorentzian linewidths for the beat-note frequency uncertainties, and sample ratios were considered and compared to the results obtained by simulating raster scans. Raster scans were not affected by linewidth uncertainty.}
\label{fig:statistics}
\end{figure}

To test the robustness and accuracy of the reconstructions with respect to noise and sample ratios, varying levels of Gaussian white noise were introduced into the simulation. Six different noise levels are presented in Fig. \ref{fig:statistics} (a). For each PSNR level, 10 different critically sampled ($m=n$) measurement vectors were generated using different sensing matrices but with the same parameters listed in section \ref{ssec:param}. Moving in increments of $2\%$, $m$ was adjusted from $.02n$ to $n$ for each measurement vector. Reconstructions were performed at each sample percentage and the non-smoothed reconstructed depth-map were compared against the true depth-map to generate a mean-squared error. Thus, every data point in Fig. \ref{fig:statistics} is the average mean-squared error of 10 reconstructions while the solid colored regions above and below each marker designate the standard error in the mean. As seen in Fig. \ref{fig:results}, the mean-squared error increases rapidly for PSNR$<5$, probably because the SNR of distant objects is well below unity. Additionally, we see that a sample ratio of roughly $m/n = .2$ is sufficient for depth-map generation at almost all noise levels. 

Within Fig. \ref{fig:statistics}, image (a) presents a reasonable way of analyzing how well all depths are recovered. However, it does not present the depth uncertainty of a single object as a function of the sample percentage and SNR. For that comparison, image (b) presents the average depth uncertainty to the toroid. Each data point was obtained by performing 10 reconstructions of 10 different data sets, finding the standard deviation in the error of each pixel (by first subtracting the true depth-map from each reconstruction), and then computing the average of the standard deviations associated with pixels comprising the toroid. These simulations were repeated with different SNR levels (using the PSNR method). Additionally, the laser-linewidth was reduced to 500 kHz and then 50 kHz (corresponding to a beat-note linewidth of 1 MHz and then 100 kHz, respectively). A simulation with an infinite PSNR and 100 kHz beat-note linewidth was included in an attempt to arrive at an upper bound on the performance when using a typical laser suited for FMCW-LiDAR. For completeness, the uncertainty in the toroid's depth from a raster scan is also included. 

Image (b) within Fig. \ref{fig:statistics} shows that the compressive depth-mapping scheme is not as accurate as raster-scanning systems, regardless of the SNR and laser linewidth -- most likely due to reconstruction errors. The compressive LiDAR camera located the toroid with an uncertainty between approximately 10 - 100 cm, depending on the SNR, linewidth, and measurement ratio, while the raster-scan consistently identified the depth within 5 cm. Because raster scanning with a FMCW-LiDAR requires finding the central frequency distribution associated with a single depth, laser linewidth and noise have little impact on identifying this value -- particularly after denoising. 

In contrast, simultaneous object detection requires isolating the peaks of different frequency components that may only be resolvable after denoising and then deconvolving the laser's linewidth. Because these operations are unlikely to be optimal (particularly in these simulations), noise and uncertainty will inevitably find their way into the measurement vectors. Additionally, the constrained optimization algorithms require a sparse basis transform $\Psi$ and user input parameters (e.g. $\alpha$) -- each of which will impact the reconstruction. Because pixels associated with a single object are likely to reside near one another, using a total-variation minimization with a small $\alpha$ parameter in lieu of the BFGS algorithm or applying a light smoothing operation within Eqs. (\ref{eq:LSxI}) and (\ref{eq:LSxInu}) will reduce the depth variation in neighboring pixels. Finally, Eq. (\ref{eq:LSxInu}) is minimized using the measurement vector $\vect{y}_{I\nu}$ -- whose elements were calculated by first using a linear frequency weighting in Eq. (\ref{eq:yvi}). As such, objects farther away contribute more to the minimization and may bias the reconstructions with nearby objects having less accuracy. Simulations suggest that moving to a sub-linear frequency weighting scheme (possibly using a square-root or logarithm) may hold better performance in identifying depths both near and far.

To summarize this section, the simulations suggest compressive depth-mapping will not be as accurate as a raster-scanned LiDAR system due a variety of tunable computational operations and basis choices. Fortunately, there is room for improvement though multiple avenues: using better denoising and deconvolution algorithms, more carefully selecting a sparse basis representation, optimizing the number of signal coefficients to operate on rather than using a fixed fraction ($m/3$), or even adding a small total-variation penalty parameter to the least-squares algorithm. Finally, trading depth accuracy for a boost in acquisition speed may not be detrimental in all applications. The speed enhancement will become even more noticeable at higher resolutions due to the increase in image compression efficiency.   

\subsection{Experimental considerations}

With the proposed experimental diagram in Fig. \ref{fig:setup} and parameters presented in section \ref{ssec:param}, an experimental realization should be straightforward. Several points to consider when building this system are presented here.

 The most difficult task is the generation of the linear frequency sweep due to the nonlinear relationship between the laser's frequency and the modulation current. Using optoelectronic feedback, this problem has been effectively solved in \cite{Satyan:09}. However, there exists a fundamental difference in a simulated versus experimental frequency sweep. The simulation used a sawtooth function for the linear sweep -- which injected aliasing artifacts into the generated frequency signals. Aliasing artifacts arose when the latter half of the delayed sawtooth interfered with the start of a new local oscillator waveform to generate high-frequency components that could not be resolved. These artifacts were merely attributed to additional noise in the simulation. However, drastic changes to a laser's modulation current can damage the laser, so it is better to use a triangle waveform. This will, in turn, inject low-frequency noise into the signal. Additionally, $1/f$ noise may be present within the system along with Gaussian white noise. Thus, a high-pass filter will be needed to clean these signals. Furthermore, the triangle function in this system may allow for compressive Doppler-mapping by using both the rise and fall of the frequency sweep.

Regarding the imaging system, the optics and detector area must be chosen such that the necessary spatial-frequency components leaving the DMD reach the difference-detector's diode. Clipping from small apertures or from insufficient detector size will lead to blurring of the depth-map.

This architecture's main limitation is in capturing quickly moving objects. The problem can be alleviated by moving to faster frame rates arising from fewer measurements. Our simulations assumed a 1 ms integration time per projection. When using a $20\%$ sampling ratio for a $128\times 128$ pixel resolution depth-map, only 3.3 sec of light-gathering time are needed. Perhaps the easiest way to reduce the acquisition time is to reduce the resolution. Reducing the resolution from $128\times 128$ pixels to $64\times 64$ pixels while maintaining the same active area on the DMD offers a factor of at least 4 timing improvement because the SNR will increase with larger super-pixel sizes. Given the short range of this LiDAR system and that some off-the-shelf DMD's can operate at 20 kHz, the 1 ms integration time can also be readily reduced by a factor of 10. 

Spatially coherent returning radiation may introduce diffraction from the DMD -- which is essentially a diffraction grating. The simulations relied on incoherent Lambertian scattering and were not affected by the DMD. Fortunately, once light has been imaged onto the DMD, the spatial profile of the light after that point is of little consequence. All that matters is that the light leaving the DMD makes it to the detector. Even if some of the radiation is lost to diffraction, a depth map can still be generated, but with a loss of SNR.

The purity of the local oscillator also plays an important role. Within section \ref{ssec:gains}, we mentioned several SNR enhancement abilities resulting from increased flux over fewer pixels, heterodyne detection, and background subtraction. There, we assumed the local oscillator contained no noise. If noise is injected into the local oscillator, many of the noise-cancellation properties disappear. Thus, it is essential that the local oscillator be phase stable with a clean frequency ramp. Fortunately, optoelectronic feedback systems used to generate the frequency sweep will also improve the phase stability of the laser \cite{Satyan:09}.

Finally, the current Python-based reconstruction algorithms take approximately 10 sec to reconstruct a $128\times 128$ depth map using a single thread on a 4.1 GHz Intel i7-4790K CPU -- preventing real-time video. Because the least-squares minimization of $\hat{\vect{x}}_{I}$ and $\hat{\vect{x}}_{I\nu}$ are independent, the reconstructions can be done in parallel. Moving the reconstruction algorithms from our serial implementation to a faster programming language, executed on a parallel architecture that takes advantage of multi-core CPUs or graphics cards, will significantly alleviate timing overhead from the reconstructions.

\section{Conclusion}

FMCW-LiDAR is a valuable technique for ranging because of its accuracy and relatively low cost. It is a natural extension to apply FMCW-LiDAR systems to depth-mapping, whether through raster scanning or other means. The compressive architecture presented in this article is an efficient means of depth-mapping that is inexpensive and can be easily implemented with off-the-shelf components. It can be easily rendered as eye-safe compared to pulsed or raster-based ranging systems. Finally, we demonstrate how to minimize the memory overhead from the sampling process. Instead of storing $m$ high-dimensional spectrograms (consisting of $m\times N$ numbers), we demonstrate how only $2m$ recorded numbers can be effectively used to recover high-dimensional depth-maps. 

This article presents an architecture to build a compressive FMCW-LiDAR depth-mapping system. Simulations support our conclusion that this system will be robust to noise when operating in the field. While there is room for improvement within the experimental design and reconstruction algorithms, we present a viable architecture that simplifies the data-taking process and significantly reduces the data storage and computation overhead. These advances serve to reduce the overall expense of such a system while simultaneously increasing the resolution and SNR for real-time depth-mapping.   

\section*{Funding}
This work was sponsored by the Air-Force Office of Scientific Research Grant No. FA9550-16-1-0359.


\end{document}